# A 4R-supported circular product–service system for luxury branded events

Ke Ma[a,c], Francesca Valsecchi[a], Yuchen Tan[a], Mingjia Ji[b], Junru Shen[a], Xiaoya Ma[a], Duan Wu[a], Jiao Mo[a], Shijian Zhao[a*]

[a] College of Design and Innovation, Tongji University, China

[b] Auditoire Asia

[c] School of Artificial Intelligence and Automation, Huazhong University of Science and Technology, China

*Corresponding author e-mail: jalab@tongji.edu.cn

**Abstract**: Temporary luxury branded events run on short cycles and bespoke builds that accelerate material churn. We present a circular phygital product–service system that operationalises the circular economy (CE) through a 4R frame (Refuse, Reduce, Reuse, and Recycling) across warehouse-to-event journeys. Developed via a multi-method design inquiry with a tier-1 contractor, the system couples physical touchpoints (reusable fold-flat transit boxes, adjustable racking, standard labels) with digital orchestration (a live digital warehouse, list-based outbound/inbound workflow, and a sustainable materials library). The architecture aligns roles and decisions, protects and identifies assets, and makes reuse the default under luxury brand constraints. By embedding traceable actions and CE-aligned rules into everyday handoffs, the PSS shifts procurement, storage, dispatch, return, and redeployment toward value retention. The contribution is a replicable, practice-ready route from circular intent to operational change in branded environments, advancing responsible retail without compromising speed or aesthetic standards.

## 1. Introduction

Short-cycle, highly customised luxury brand events have become central vehicles for storytelling and sales activation. In this paper, short-cycle or customised brand events refer to temporary branded environments—such as pop-ups, HPP booths, activation spaces, and short-duration showcase events—whose design, fabrication, deployment, and strike occur within compressed project windows, often ranging from days to a few weeks. However, the rapid turnover of materials associated with these events presents a significant tension with the principles of the circular economy. Evidence across fashion and textiles highlights the substantial environmental impacts and the need to enhance value retention through product longevity, reuse, and high-quality recovery, as opposed to promoting linear consumption

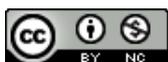

(Ellen MacArthur Foundation, 2017; Geissdoerfer et al., 2017; Ghisellini et al., 2016; Kirchherr et al., 2017; Korhonen et al., 2018; Niinimäki et al., 2020). Moreover, CE literature increasingly calls for coordinated, system-level strategies that reorganise practices, infrastructures, and information flows across supply chains and actor networks (Blomsma & Brennan, 2017; Ceschin & Gaziulusoy, 2016; Ki et al., 2020; Moorhouse & Moorhouse, 2017; Reike et al., 2022).

Within this landscape, the impact of temporary branded spaces such as pop-ups and high-profile promotions (HPPs) is especially negative: short lifespans, bespoke aesthetics, and rapid turnarounds amplify material and logistical footprints. We focus on luxury fashion and beauty events because these settings intensify aesthetic specificity, quality control, and time pressure, making circularity both more difficult and more revealing as a design challenge. A recent life cycle assessment (LCA) of a luxury fashion brand's HPP event booth quantifies substantial daily impacts relative to permanent retail builds. It identifies material hotspots that are highly sensitive to design choices, directly motivating system-level opportunities beyond one-off design swaps (Wu et al., 2025). Methodologically, LCA frameworks (ISO 14040/14044) structure how upstream design decisions propagate into downstream impacts across a booth's condensed yet intensive life cycle, making them an appropriate evidence base for design intervention (ISO, 2006a; ISO, 2006b; see also Goedkoop & Spriensma, 2001). To translate these insights into practice, we draw on Product–Service System (PSS) literature on foundational definitions, business-model implications, and operational coordination across actors and touchpoints (Annarelli et al., 2016; Baines et al., 2007; Bocken et al., 2016; Kjaer et al., 2019; Mont, 2002; Reim et al., 2015; Tukker, 2015). Prior design studies have also shown that PSS approaches can reorganise complex service settings, such as in-flight food systems and equine air transport, by aligning products, interfaces, stakeholders, and backstage operations into a single integrated offering (Trapani et al., 2022, 2025). In parallel, digitalisation within PSS, referred to as "PSS 4.0," demonstrates how data capture and workflow integration can enable circular practices at scale (Bressanelli et al., 2018). In retail and hybrid retail contexts, service design scholars similarly show how phygital journeys, role allocations, and backstage processes influence circular outcomes; these steps are critical for moving CE targets from rhetoric to practice in store/event systems (Grootboom et al., 2024). However, the back-end of branded events, including warehouse intake, storage, dispatch, return, and redeployment, remains under-specified in the CE and PSS literature, even though these flows are decisive for closing loops under time pressure and brand-quality constraints. Broader CE syntheses claim that value retention depends on concrete operational schemas and business model choices that genuinely capture the intended environmental value propositions (Lieder & Rashid, 2016; Lüdeke-Freund et al., 2019; Manninen et al., 2018; Reike et al., 2018). For short-term, high-value event materials, this implies tightly coordinated processes and traceable information that enable reuse at speed and accountability by design.

This paper responds by proposing and documenting a circular PSS that links back-end warehouse operations to front-stage brand workflows for luxury-brand events, and by outlining an implementation strategy for an environmental value proposition. Specifically, we articulate a phygital intervention comprising: (i) physical touchpoints—reusable transit boxes and adjustable racks that enable design-for-disassembly and rapid redeployment; (ii) digital touchpoints—a sustainable digital platform with three core modules; and (iii) measurement constructs derived from event-process data to compute circular, operational, service-quality, and carbon indicators. We treat the event system end-to-end, so reuse becomes the default rather than the exception while remaining compatible with luxury-brand requirements and standards.

## 2. Literature review

### 2.1 CE in Fashion and Events

The CE reframes value retention around slowing, narrowing, and closing resource loops rather than linear throughput (Blomsma & Brennan, 2017; Geissdoerfer et al., 2017). Fashion has been a focal sector due to documented environmental burdens across extraction, production, consumption, and disposal (Niinimäki et al., 2020; Moorhouse & Moorhouse, 2017). Foundational reviews map CE drivers, strategies, and system boundaries while also surfacing conceptual heterogeneity across the industry and policy–practice gaps and shortfalls (Ghisellini et al., 2016; Kirchherr et al., 2017; Korhonen et al., 2018; Reike et al., 2018, 2022; Bolzan et al., 2024). For textiles and apparel, industry reports, and academic literature emphasise durability, reuse, and infrastructure for traceability and reverse flows as preconditions for meaningful impact (Ellen MacArthur Foundation, 2017; Lieder & Rashid, 2016; Manninen et al., 2018; Lüdeke-Freund et al., 2019). Within this milieu, temporary branded environments such as pop-ups and HPPs combine bespoke aesthetics with fast build–strike cycles, amplifying material churn and logistics intensity relative to permanent stores.

### 2.2 Product–service systems for circularity

PSS literature positions service logic and coordinated product–service architectures as levers to decouple delivered value from material throughput (Mont, 2002; Manzini & Vezzoli, 2003; Baines et al., 2007; Tukker, 2015). Systematic reviews consolidate PSS definitions, taxonomies, and business model tactics pertinent to CE aims (Annarelli et al., 2016; Reim et al., 2015; Kjaer et al., 2019), while design-for-sustainability work traces the evolution from product-level eco-design to socio-technical transitions and system innovations (Ceschin & Gaziulusoy, 2016). Empirical cases of digitalised PSS ("PSS 4.0") demonstrate how sensorisation, identification, and platform orchestration enable condition monitoring, asset tracking, and value-retention loops in practice (Bressanelli et al., 2018). In the fashion industry specifically, stakeholder engagement is necessary to align brand, operations, and supply chain actors around CE targets (Ki et al., 2020).

### 2.3 Phygital service design and retail event systems

Service design research highlights how coordinated physical artefacts and digital interfaces shape behaviours across journeys, backstage processes, and adaptive interactions (Grootboom et al., 2024; Ma & Cao, 2019; Shi & Ma, 2018). Related studies of place-based and multisensory experience systems show that integrated online-offline strategies and staged experiential media influence how people perceive and move through designed environments (Jiang et al., 2025a; Valsecchi et al., 2024). Together, CE explains why value retention matters, PSS explains how value can be reorganised across actors and offerings, and phygital service design explains how touchpoints and information visibility successfully enact circular intent.

*2.4 Standards and measurement frameworks*

LCA offers a comparative basis for judging design and logistics options when aligned with recognised standards, while damage-oriented methods, such as Eco-indicator 99, translate impacts into pathways relevant to HPP material and architectural choices (ISO, 2006a; ISO, 2006b; Goedkoop & Spriensma, 2001; Wu et al., 2025). We treat data capture itself as a design task, deciding what to log, when, and by whom. In this way, there is consistent tracking of reuse rates, cycle times, loss or damage, and condition grading, linking environmental accounting with everyday warehouse decisions (Kjaer et al., 2019; Manninen et al., 2018).

*2.5 Problem statement and research gap*

Despite advances in CE frameworks for fashion and the maturation of PSS theory and practice, the back-end warehouse operations for event-booth materials remain under-specified: returns, inspection, sorting, reconditioning, and rapid redeployment are rarely modelled as an integrated, phygital service system tailored to HPP constraints (Kirchherr et al., 2017; Korhonen et al., 2018; Kjaer et al., 2019; Grootboom et al., 2024). Existing CE and PSS literature establishes the why and the what, and LCA clarifies the where. The remaining challenge is how to operationalise a context-specific phygital architecture that enables rapid reuse and makes traceability the default in luxury HPP workflows.

Accordingly, this study asks three research questions: RQ1: How can a circular PSS connect back-end warehouse operations with front-stage luxury event workflows? RQ2: Which physical and digital touchpoints are required to operationalise the 4R frame under conditions of speed, aesthetic control, and traceability? RQ3: What early operational and circularity effects emerge when such a phygital PSS is deployed in practice?

## 3. Research methods

We investigated a real-world case involving a Tier-1 brand-events contractor in China (Auditoire), part of a global brand-experience group that delivers more than 50 events per year for multiple luxury fashion houses and operates a central warehouse. Auditoire was selected as a revelatory single case because it concentrates the tensions addressed here: repeated short-cycle delivery, a shared warehouse, multiple stakeholder groups, and strict luxury-quality standards. We therefore adopted a staged, multi-method design-research strategy with clear sequencing and feedback loops: workshops (April 2024), observations and survey (May 2024), interviews and document audit (June 2024), followed by synthesis, prototyping, and staged deployment.

*3.1 Data collection*

We began with three stakeholder workshops in April 2024. Participants included the research team's PSS, product, and interaction designers, as well as Auditoire's project managers, warehouse administrators, procurement staff, materials specialists, and the sustainability lead. Each session lasted approximately ninety minutes. The first workshop clarified responsibilities, required information, interdependencies, and recurrent tensions among roles. The second workshop reconstructed how a single item moves across processes, for example, how materials are counted and brought back into the warehouse after an event, together with who is accountable at each step and which communication channels they rely on. The third workshop examined the current warehouse space and ways of managing items, documenting disorder and articulating expectations for storage and retrieval. Outputs were consolidated into a stakeholder map (Figure 1) and provided the initial scaffolding for a service blueprint.

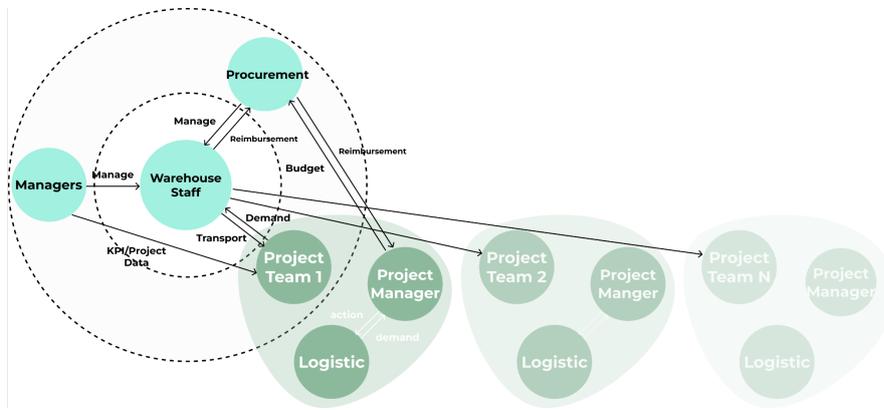

*Figure 1. Stakeholder map.*

In May 2024, we conducted on-site non-participant observations at the warehouse, where we observed how items are stored and stacked; how teams select and pack items for outgoing projects; how records are created; how shipments are prepared; and how items are dismantled, counted, and re-shelved when projects conclude. Particular attention was paid to day-to-day rhythms, typical causes of losses, and sources of inventory discrepancies. The observation was reiterated at the temporary storage and backstage area during HPP preparation, where we observed how materials are staged, controlled, and later counted for return. All observations were conducted with prior authorisation and documented with photographs and short videos. These records were used to refine the service blueprint, confirm the order of operations, and locate critical touchpoints (Figure 2).

To probe opportunities for a circular PSS and to prioritise issues emerging from the field, we ran a targeted survey in May 2024. The instrument comprised 20 questions covering awareness of the CE and sustainable design, current workflows, client expectations of luxury and fashion brands regarding sustainability, perceived difficulties, and suggestions for improvement. Respondents were purposively selected because they directly participated in event planning, materials specification, procurement, warehouse handling, client coordination, or sustainability oversight. We collected 37 valid responses across the design, technical, customer service, lifestyle, and management departments. The survey identified the most widespread frictions in the existing system and informed the prioritisation of touchpoints addressed in RQ2.

In June 2024, we conducted 9 semi-structured interviews with internal staff: 2 warehouse administrators, 2 project managers, 1 sustainability lead, 2 designers, 1 procurement staff member, and 1 finance reviewer. These roles were selected to cover the main decision points and handoffs across the warehouse-to-event journey. Interviews were tailored from the survey's focal issues. They investigated the causes of problems and pain points, including system-level constraints, execution variations, limitations of communication tools, and challenges in cross-departmental coordination. Each interview lasted about one hour and, with permission, was audio-recorded and documented in notes.

Finally, we assembled and examined artefacts used across process nodes, including operational forms, records, checklists, chat transcripts, and warehouse archiving materials. These sources were used to complete the information architecture and to verify key fields and handoffs identified through workshops, observations, the survey, and interviews. The five methods mutually corroborated one another and created a traceable evidence base from which the stakeholder map, the service blueprint, and subsequent design requirements for the phygital system were derived.

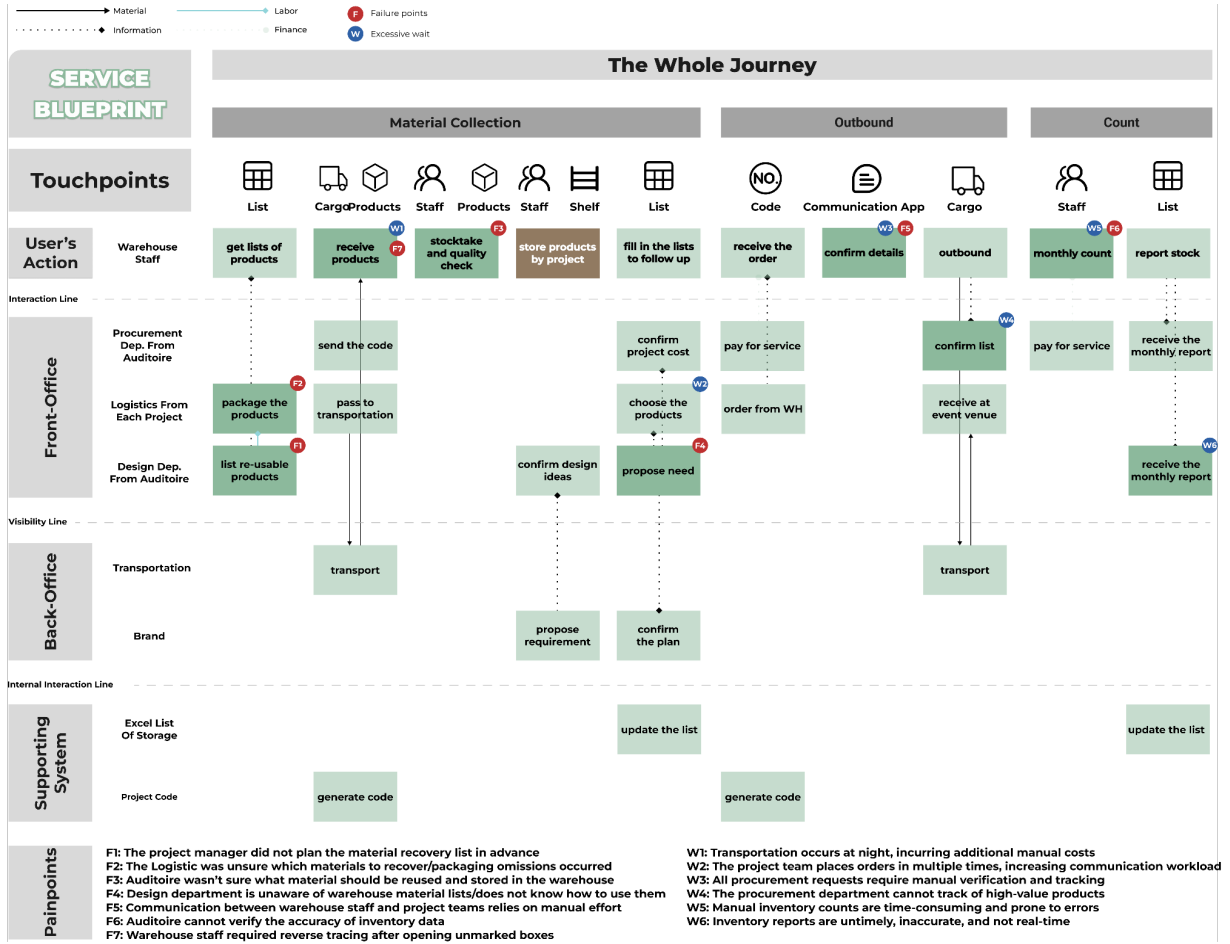

*Figure 2. Service blueprint.*

## 3.2 Data analysis

Our analysis followed the study's staged logic, moving from holistic system visualisation to problem identification and, finally, to concept development and early prototyping. Each step drew on the specific evidence produced by the five methods and was designed to convert heterogeneous observations into designable requirements for a circular PSS.
First, we visualised the service system. Insights from the three stakeholder workshops were consolidated into a stakeholder map (Figure 1) that clarifies the actors, their responsibilities, the required information, and recurrent tensions across the end-to-end journey. We then integrated records from on-site non-participant observations and the interviews to construct a service blueprint (Figure 2) that traces the current sequence of activities from initial requirement and approval through picking, dispatch, return, and inspection. The blueprint details role handovers, interactions with materials at each step, and the physical and digital touchpoints used, providing a standard reference model for subsequent analysis.
Second, we identified problems and service gaps at three levels. At the system level, the targeted survey of thirty-seven employees was analysed to prioritise sustainability-relevant issues and opportunities. We report the ten most urgent problems (Table 1) and the five most frequently endorsed opportunity areas (Table 2). At the process level, the nine interviews were coded by role and scenario to locate pain points at critical stages of the journey; results are presented as a cross-role view of breakdowns and candidate interventions (Table 3). At the touchpoint level, evidence from warehouse observations and interviews with administrators showed that non-standard cardboard boxes and ad hoc stacking led to mixed material conditions, difficult retrieval, and inconsistent record-keeping. Photos illustrate the disorder and the effort required to stabilise inventory states (Figure 3).

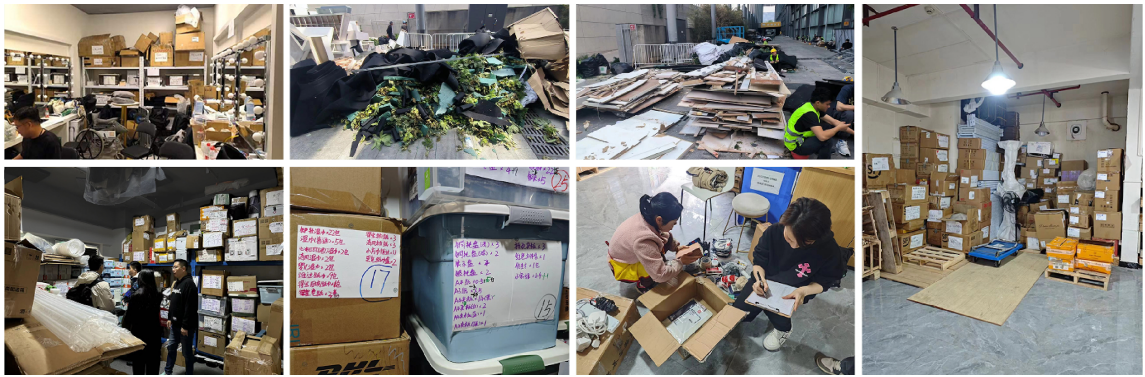

*Figure 3: On-site non-participant observations of the used warehouse and backstage spaces.*

Third, we developed concepts and produced early prototypes. Synthesising findings from the stakeholder map, service blueprint, survey, interviews, and field records, the team articulated three design insights that structure the phygital intervention.
1. Material storage and handling are a primary source of loss, re-purchasing, and avoidable waste. To address this, we specified physical touchpoints that enable durable protection and standardised organisation—reusable transit boxes and adjustable racks—so that items can be located, counted, and redeployed with less effort and lower risk.
2. Even when physical organisation improves, information about what is available, where it is, and in what condition remains opaque to the broader team. We therefore defined a digital warehouse that publishes transparent, dynamically updated inventory states to all authorised stakeholders, reducing one-to-one gatekeeping and bottlenecks.
3. With inventory visible, coordination still depends on fragmented messages and manual confirmations, which weakens accountability and prevents consistent measurement. We therefore extended the platform to orchestrate the flow of materials and decisions across projects, with role-based permissions and traceable actions, to enable the routine capture of operational and CE-aligned indicators, such as reuse rate, cycle time, loss and damage, and inventory accuracy, without imposing additional pressure on frontline teams.
Concepts were advanced through iterative prototyping. Physical touchpoints were mocked up and tested in the warehouse to assess basic fit, handling, and stacking behaviour; feedback from administrators informed revisions to dimensions, labelling surfaces, and consolidation practices. In parallel, a rapid digital prototype of the digital warehouse and role-aware operations, as the platform's core services, was shown to stakeholders to refine metadata, flows, permissions, and required modes of interaction. These cycles established the specifications for the integrated phygital system presented in the following section.

*Table 1: Top ten urgent sustainability problems (multi-select; n = 37)*

| Rank | Problem | Votes | Share |
|---|---|---|---|
| 1 | Excess use of non-recyclable materials in the event build | 29 | 78.38% |
| 2 | Storage and management of materials and products | 18 | 48.65% |
| 3 | Prevalence of single-use materials | 16 | 43.24% |
| 4 | Overproduction of printed materials beyond actual needs | 15 | 40.54% |
| 5 | Lack of event LCA | 15 | 40.54% |
| 6 | Difficulty selecting certified sustainable materials | 14 | 37.84% |
| 7 | Communication with clients on sustainability | 13 | 35.14% |
| 8 | On-site waste awareness and waste management | 13 | 35.14% |
| 9 | One-off purchase and use of technical equipment | 11 | 29.73% |
| 10 | Transport carbon footprint of materials and products | 11 | 29.73% |

*Table 2: Opportunity areas — feasibility summary (n = 37)*

| Rank | Opportunity area | High-feasibility votes | Low-feasibility votes |
|---|---|---|---|
| 1 | Sustainable materials library to guide specification | 36 | 1 |
| 2 | Optimisation of storage and logistics structures | 34 | 3 |
| 3 | Secondary and multiple reuses of post-use materials | 34 | 3 |
| 4 | Standardised recovery process and shared database | 33 | 4 |
| 5 | Multi-party collaboration platform for stakeholders | 21 | 6 |

*Table 3: Stakeholder-specific pain points*

| Category | Warehouse staff | Project managers | Design team | Finance and procurement |
|---|---|---|---|---|
| Physical storage and standardisation | Non-standard cartons and no coding make picking slow and increase the risk of damage. | Physical stock is disorganised, so availability cannot be confirmed in advance. | Sample check-outs and item condition are unclear, delaying specification. | Low space utilisation raises carrying costs and leaves assets idle. |
| Digital transparency and traceability | Inventory maintained in Excel—inefficient and error-prone; hard to | No real-time view of inventory and reservations, leading | Cannot search reusable materials by attributes and | No traceability or visibility of cost versus recovered |

|  | answer queries. | to conflicts and rescheduling. | current availability. | value across projects. |
|---|---|---|---|---|
| Process orchestration and cross-role coordination | Requests are fragmented and lack a unified picking cadence, resulting in heavy one-to-one coordination. | Approvals and handoffs are scattered across chats with no transparent responsibility chain or service levels. | Design milestones are misaligned with the warehouse gates, leading to last-minute substitutions. | Approval rules for high-value items are applied inconsistently. |

## 4. Results

Building on three months of research, analysis, design, and early prototyping, we transitioned to systematic development, application, and formative evaluation of the circular PSS. Our results are presented from three complementary angles.

### *4.1 CE practices*

Grounded in system-level contradictions and stakeholder pain points, the team articulated a context-specific "4R" frame—Refuse, Reduce, Reuse, and Recycling—and mapped where each practice should intervene across the service journey. The resulting system map (Figure 4) clarifies how these practices operate in combination rather than isolation, strengthening circular performance without undermining speed or aesthetic standards.

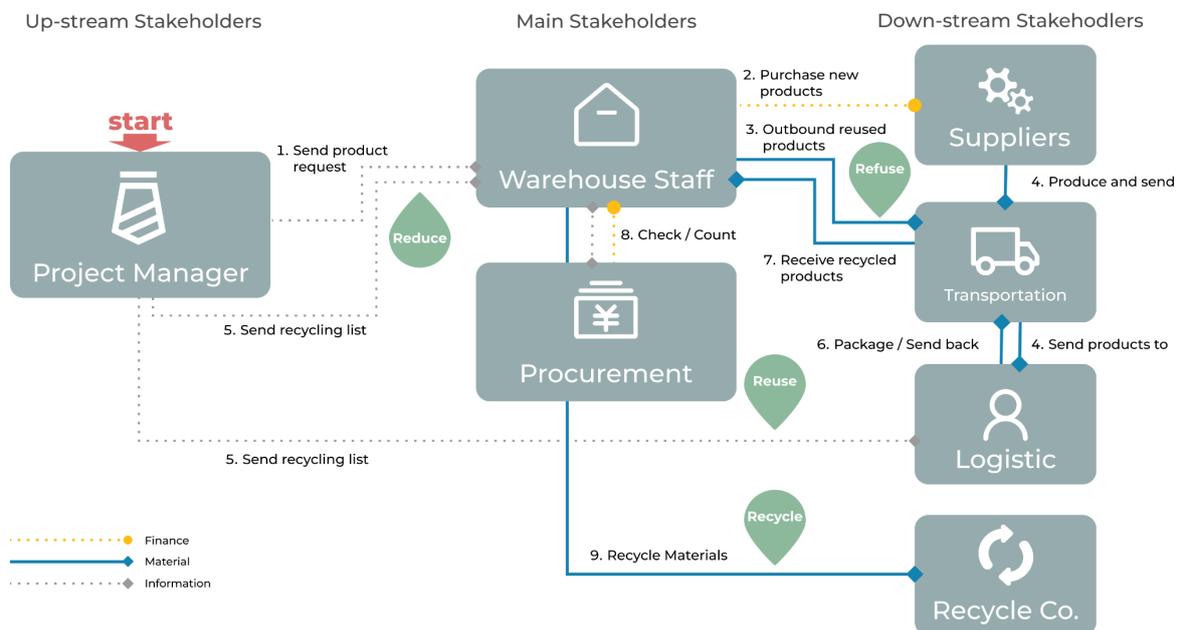

*Figure 4. System map illustrating the new PSS applying the 4R-frame.*

**Refuse** denotes avoiding unnecessary purchasing when suitable items already exist. With the digital warehouse making inventory states visible and searchable, designers and project managers can identify viable substitutes from current stock, and procurement can align requests with what is already available, thereby rejecting new buys that do not add value.

**Reduce** focuses on cutting avoidable loss and waste during live projects. Transparent project status and precise picking and return lists lower the incidence of misplaced items, partial returns, and over-ordering. By aligning responsibilities and timing, the platform helps shrink idle stock and redundant movements.

**Reuse** targets systematic redeployment of materials across projects. Metadata such as quantity on hand, condition grading, and remaining lifespan support confident specification of items for the next build. In parallel, physical touchpoints—reusable transit boxes, adjustable racks, and standardised labels—protect items, preserve identity, and accelerate reconfiguration between cycles.

**Recycling** addresses end-of-use routing when items no longer meet reuse thresholds. Upon return, operators assess the condition and allocate items to traceable recovery streams, with the platform recording these decisions so that material recovery is auditable and tracked against future specifications.

Guided by this 4R frame, we specified the corresponding physical and digital touchpoints and instrumented data capture for CE- and operations-relevant indicators. This enables ongoing calculation of reuse rates, cycle times, losses and damage, and inventory accuracy as the system is deployed and refined.

## 4.2 Touchpoints

### 4.2.1 Physical touchpoints

Our physical touchpoints concentrate on the warehouse space, addressing concrete issues of storage, handling, and visibility so that material information becomes transparent and traceable, and reuse becomes efficient. By standardising containers and racking, stabilising how items are packed and retrieved, and making states legible at a glance, the physical layer underpins the accuracy and timeliness of the digital warehouse as projects progress.

**Reusable Transit Boxes.** Day-to-day operations relied on ad hoc corrugated cartons, frequent re-boxing, and improvised labelling that mixed states and obscured provenance. We introduced a family of reusable, fold-flat transit boxes that circulate from storage to dispatch and return without repacking. Two interlocking dividers split the interior along both axes to stabilise partial loads and support fast consolidation (Figure 5). Boxes are fabricated from polypropylene honeycomb board for a high stiffness-to-weight ratio, abrasion resistance, and long service life, with scored folds and reinforced corners to withstand repeated handling. In initial trials, the boxes reduced reliance on improvised packaging, improved the segregation of new and returned items, and provided a stable physical anchor for inventory identity in the digital warehouse (Figure 6).

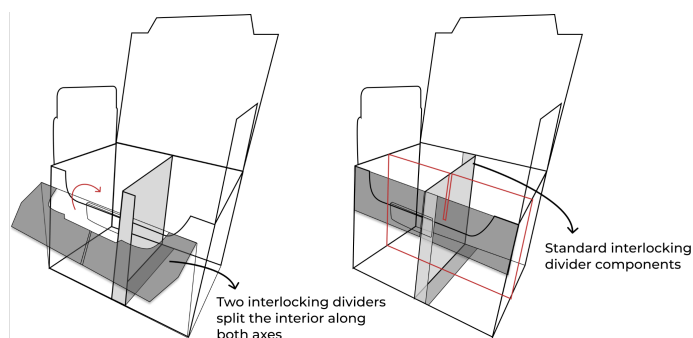

*Figure 5. Transit-box structure diagram.*

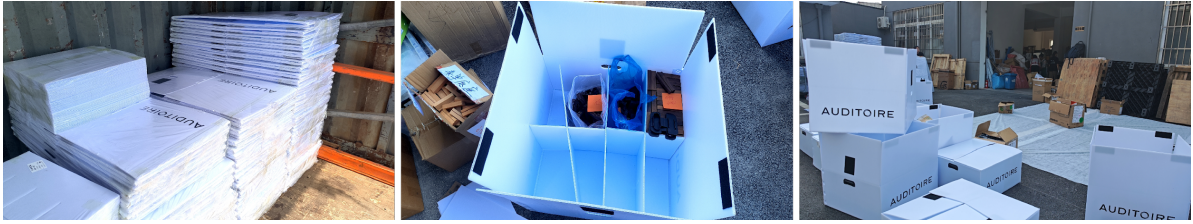

*Figure 6. The manufactured reusable transit boxes are used in the warehouse.*

**Adjustable Racking.** Non-standard cartons created dead space, unsafe piles, and congested aisles; picking relied on memory, and packing occurred away from storage. We deployed adjustable racking sized to the transit-box footprint to remove voids and relieve aisle congestion: shelf pitch can be repositioned to accommodate mixed box heights, and rack bays can be compacted or opened via floor rails with lockable castors. The addition of clip-in modules, composed of a small work surface and a hanging rail, brings packing, photography, and labelling tasks directly into the rack zone (Figure 7). The frames use aluminium extrusions with standard corner connectors; a gridded acrylic deck set on four rotatable supports creates a stable, reconfigurable storage plane. The prototype bay delivered clearer sightlines, improved cube utilisation, safer picking, and fewer temporary stacks, aligning the physical topology with the digital record (Figure 8).

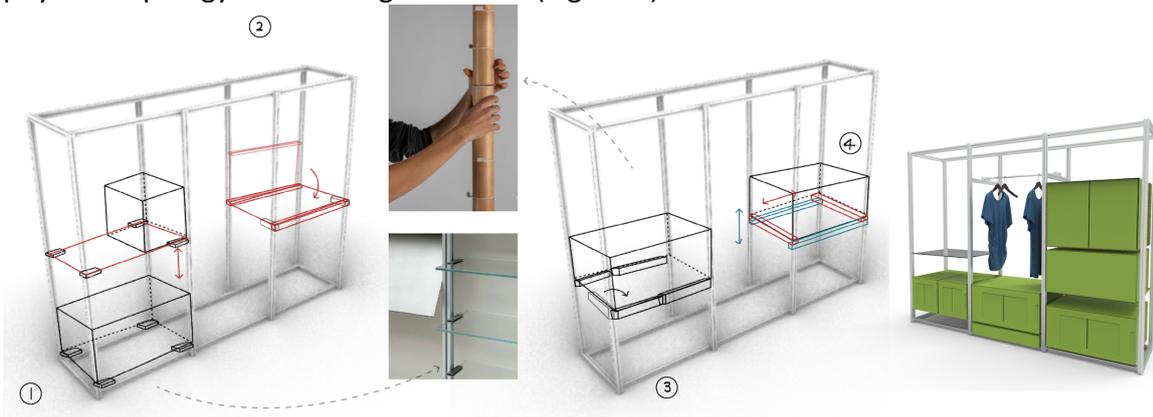

*Figure 7. Adjustable racking principle's diagram.*

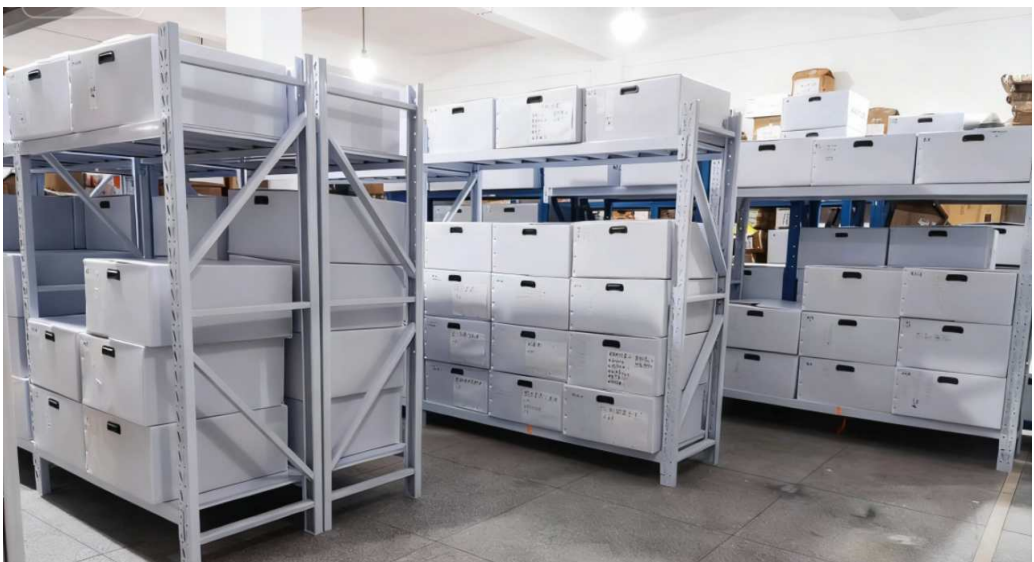

*Figure 8. Racking in use.*

### 4.2.2 Digital touchpoints

Our sustainable digital platform has three modules that work together to support the 4R frame. They make material states visible, guide hand-offs across roles, and record every decision so that refusing unnecessary purchases, reducing loss, reusing assets, and routing items for recycling become routine rather than exceptional.

**Digital warehouse.** The digital warehouse is the backbone. It keeps a live list of all items with search and filter, and an item page that shows the essentials: stock quantity, condition, material, expiry date, indicative carbon footprint, and status. Each item carries a unique label. Records are updated whenever teams pick, pack, dispatch, receive, inspect, or return items, keeping stock levels accurate and movements traceable across projects (Figure 9). The digital warehouse currently holds 1,380 distinct items across the seven main categories of event props, medical supplies, electronics and electrical equipment, office supplies, beverages and food, apparel and footwear, and tableware and glassware. Warehouse administrators can upload new items, delete obsolete items, and adjust stock quantities and metadata to keep inventory information dynamically up to date.

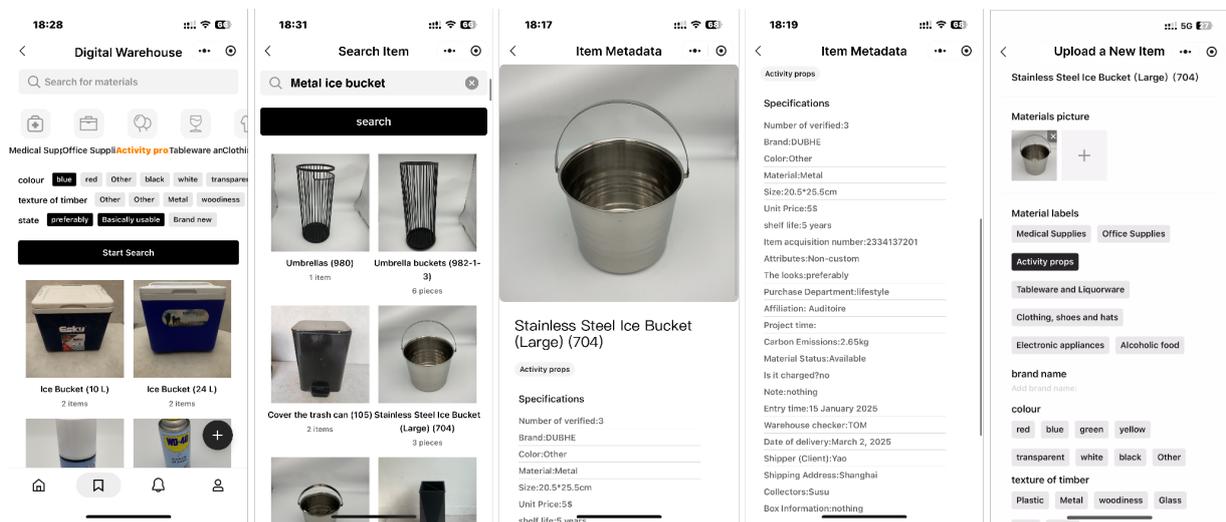

*Figure 9. Selected digital warehouse screens.*

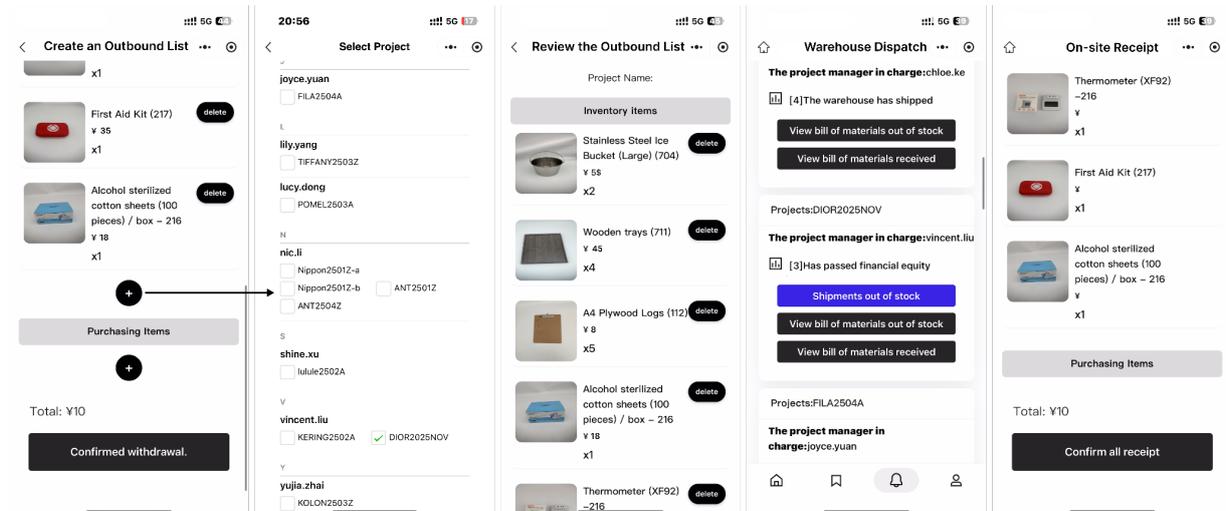

*Figure 10. Selected project workflow screens centred on the outbound list, with role-based actions, and milestone confirmations.*

**Project logistics workflow.** Project work is organised around outbound and inbound lists. A project lead creates an outbound list by selecting items from the digital warehouse. That list then guides approvals, picking, packing, dispatch, and on-site receipt. When the event ends, the same list becomes an inbound list. Each line is marked as returned for restocking, consumed or damaged, or temporarily stored for later use. This handover keeps items from dropping out of sight and ties every on-site outcome to a return decision. At each milestone of the item's journey, role-specific notifications prompt the next person to act. Permissions ensure that only authorised roles can confirm steps that change the inventory state. Once the inbound list is reconciled, stock updates immediately in the digital warehouse, and the project dashboard reports cycle time, reuse rate, loss, and related indicators (Figure 10).

Sustainable materials library. The materials library helps designers choose reusable, lower-impact options for booths, fixtures, and displays, aligned with each project's outbound list and build plan. Designers browse by category and property, view material details and guidance, and use an AI agent to query the local materials knowledge base and receive suggestions. Selected materials link back to the project so choices flow through to picking and return steps without extra work (Figure 11). In this sense, the library combines structured material intelligence with AI-assisted design support, echoing work on sustainable service ideation, human meaning-making, and meta-designed agent interfaces (Ge et al., 2026; Jiang et al., 2025b; Lou, 2023).

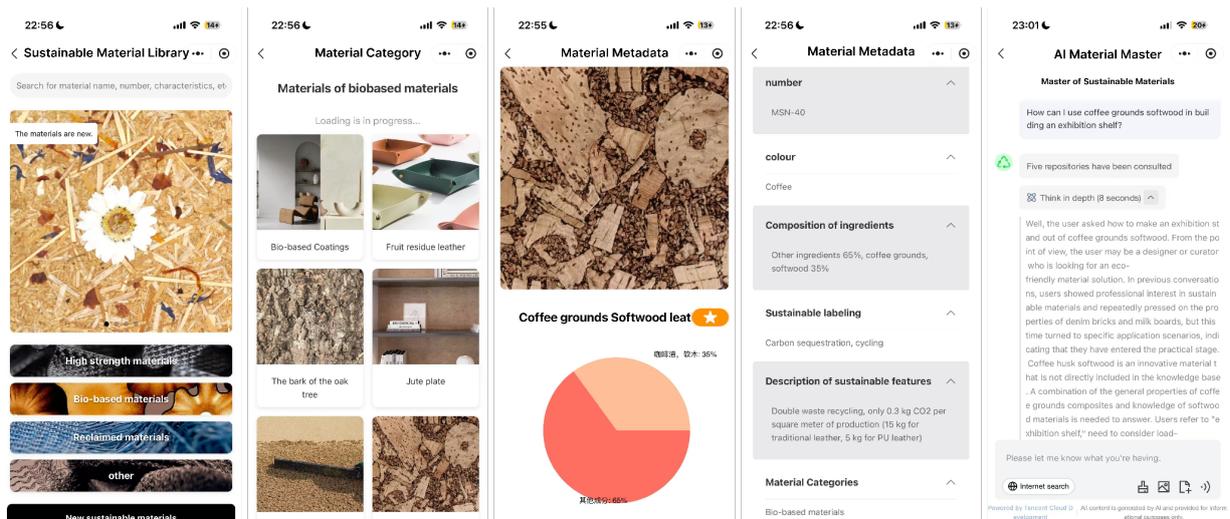

*Figure 11. Selected sustainable materials library screens.*

### 4.3 Activities

After three months of research, analysis, and prototyping, we moved into staged development focused on touchpoint delivery and deployment. Phase 1 (November 2024) established the digital warehouse and aligned it with the new physical touchpoints (reusable transit boxes and adjustable racks); internal rollout began immediately. Phase 2 (April 2025) released the project logistics workflow, moving from pilot to regular use across live projects. Phase 3 (June 2025) launched the sustainable materials library, integrated with ongoing projects, so designer selections flowed directly into the platform.

By the time of writing, 22 projects have run end-to-end on the system, confirming that the three touchpoints are in steady use and supporting day-to-day operations. Here, we present one representative short-cycle project to illustrate system operation, scope, timeline, and indicative LCA dashboard outputs. Although the client was a global technology firm rather than a luxury fashion house, the project exercised the same warehouse, material, and coordination logic developed for luxury-branded event operations. It therefore served as a live stress test of the proposed PSS. Auditoire delivered multiple interactive zones, namely a "future phone booth," an Android robot experience, a graffiti studio, and AI rapid-build workshops, and used the circular PSS to coordinate planning, warehouse operations, and on-site logistics. All fixtures, furniture, and props were sourced from existing warehouse stock rather than purchased new. Hence, the project-level outstore record contains only outbound entries and no purchase lines, manifesting the Refuse dimension of the 4R frame. Across all categories, the outstore list shows 394 individual item units dispatched from the warehouse. In the corresponding inbound record, 190 units were explicitly marked by the team as consumables, 198 units were returned and restocked, and 6 units were left temporarily at the venue for continued use. If we use the 204 items intended for reuse as the denominator, this yields an effective recovery rate of approximately 97%, with only a small number of items pending transfer back to the warehouse.

Using ISO 14067 product carbon footprint principles and generic emission factors for wood-based props, metal fixtures, and plastic elements, we approximated the embodied emissions of the 198 returned and restocked items by category. This conservative calculation suggests that reusing these items instead of procuring equivalent new stock avoids roughly 3 t $CO_2$ for this project. This pattern illustrates how combining physical touchpoints and list-based digital workflows can support high recovery of reusable assets in a complex, short-cycle event without additional purchases.

## 5. Evaluation

We ran a post-deployment survey in October 2025 with 14 respondents who had directly used the system across 22 live projects (7 project managers, 5 warehouse administrators, and 2 sustainability leads). The survey investigated which problem areas were addressed, captured performance on five indicators using two complementary scales, and assessed which CE concepts were enacted in day-to-day work.

Respondents most often selected material reuse (14 of 14, 100%) and process transparency (12 of 14, 85.7%). Cycle-time reduction and error reduction were each selected by 10 of 14 (71.4%). Traceability and warehouse orderliness were each selected by 9 of 14 (64.3%). The lead position on reuse and transparency aligns with the design intent to make existing inventory visible before purchasing decisions and to standardise handling so stored items can be redeployed with confidence.

We report, for five indicators, both an absolute post-deployment score from 0 to 100 and an improvement ratio derived from a 1–5 improvement rating (mean divided by 5). Indicators were process transparency, operation time, Information accuracy, warehouse orderliness, and material reuse rate. Transparency and reuse show the most substantial gains, with orderliness and time close behind. Accuracy improves meaningfully, but dispersion across teams remains greater.

When asked which CE concepts the circular PSS helped implement, responses focused on the 4Rs: Reuse (100%), Reduce (85.7%), Recycle (78.6%), and Refuse (71.4%). Additional concepts received lower endorsements: Repair (28.6%), Refurbishment (21.4%), Remanufacture (21.4%), and Recovery (14.3%). This pattern reflects a first wave of gains, in which visibility and control over existing inventory most directly affect day-to-day choices.

Findings (Table 4) indicate that the combination of a digital warehouse, list-based project logistics, reusable transit boxes, and adjustable racks shifted everyday practice toward reuse and transparency, shortened coordination loops, improved spatial order in the warehouse, and raised task accuracy. The improvements align with the research ambitions. Workshops and observations revealed where visibility and hand-offs failed. The survey and interviews prioritised issues that phygital touchpoints could address. The resulting system made inventory states visible to all roles, enforced standard labels and list progression, and protected materials physically, resulting in measurable gains on the targeted indicators.

*Table 4 Indicator performance (n = 14).*

| Indicator | Absolute score (0–100) | Improvement ratio |
| --- | --- | --- |
| Process transparency | 82 | 0.88 |
| Operation time | 78 | 0.82 |
| Information Accuracy | 74 | 0.78 |
| Warehouse orderliness | 80 | 0.84 |
| Material reuse rate | 79 | 0.86 |

## 6. Discussion and conclusions

This study treated luxury branded events as an end-to-end service system and yielded two lessons relevant to design. First, circular performance is secured less at the customer-facing moment than in backstage decisions about storage, visibility, condition grading, approvals, and return pathways. Second, information behaves like a material: when item identity, condition, location, and responsibility are traceable, reuse becomes feasible rather than aspirational. The findings extend circular PSS research by showing that value retention in temporary branded environments depends on orchestrating physical assets and informational handoffs, and they shift service design attention from front-stage circular experiences to backstage infrastructures where circular outcomes are decided.

Two limitations limited the value of the outcomes and indicated where the design must work harder. Personnel adaptation lagged behind the new PSS, with shadow practices, partial rollout, and heterogeneous bespoke items creating gaps between specification and reality. In parallel, digital–physical drift and only partial LCA integration constrained traceability and environmental claims, as missed scans or delayed updates desynchronised records from shelves, and carbon linkages remained scenario-based rather than fully automated.

Future work should test the 4R-guided PSS across more event formats and client governance conditions. The orchestration logic may also transfer to other high-turnover temporary service systems, including exhibitions, trade fairs, hospitality installations, and cultural activations.

In conclusion, this paper advances a 4R-guided circular product–service system that reorganises warehouse-to-event flows so that refusing avoidable purchases, reducing loss, reusing assets at speed, and routing residuals for recycling become routine under luxury constraints. The core move is orchestration: a shared logistics spine, role-aware gates, and physical identity carriers that keep information attached to materials. By treating the warehouse as a primary site of design and aligning physical and informational touchpoints, the work demonstrates a designerly route from circular intent to operational change in branded environments.

**Acknowledgements:** This work received support from New Generation Artificial Intelligence-National Science and Technology Major Project (2025ZD0122801); the National Natural Science Foundation of China under Grants 62276063 and U23B2057; Research for Child-Friendly Healthcare Experience Design (22YJA760054); and Future Disciplines of Shanghai Municipal Education Commission: Construction in Extreme Environments.

About the authors:

**Ke Ma** is affiliated with the College of Design and Innovation, Tongji University, and the School of AI and Automation, Huazhong University of Science and Technology. He contributed to the circular PSS, digital touchpoints, and AI-enabled design methods for complex service environments.

**Francesca Valsecchi** is an Associate Professor at the College of Design and Innovation, Tongji University, China. Her research and teaching explore systemic design methods and tools, sustainability and design scope, and human-nature interactions. She contributed to the framing of circularity, phygital orchestration, and the implications for design research.

**Yuchen Tan** is affiliated with the College of Design and Innovation, Tongji University, China. In this project, Yuchen contributed to the investigation of stakeholder journeys, service blueprints, and the development of digital touchpoints for circular warehouse-to-event coordination.

**Mingjia Ji** is affiliated with Auditoire Asia. His professional practice centres on branded sustainability design and event delivery. He is the project lead representing Auditoire Asia to collaborate with the Tongji research team.

**Junru Shen** is affiliated with the College of Design and Innovation, Tongji University, China. His contribution to this study lies in service-system analysis, prototyping, and the integration of physical and digital touchpoints.

**Xiaoya Ma** is affiliated with the College of Design and Innovation, Tongji University, China. In this research, Xiaoya contributed to the design inquiry and the development of user-facing elements that connect warehouse information and material selection.

**Duan Wu** is affiliated with the College of Design and Innovation, Tongji University, China. Her research interests include sustainability assessment, circular economy, and environmental measurement.

**Jiao Mo** is an Associate Professor at the College of Design and Innovation, Tongji University, China. Her work focuses on design innovation, systems thinking, and product–service system design.

**Shijian Zhao** is the director of the Jewellery Lab at the College of Design and Innovation, Tongji University. Her research addresses strategic design and sustainable innovation, benefiting from her guidance in aligning circular design methods with organisational practice and feasibility.